# The KR-Benes Network: A Control-Optimal Rearrangeable Permutation Network

Rajgopal Kannan
Department of Computer Science
Louisiana State University
Baton Rouge, LA 70803
email: rkannan@csc.lsu.edu

**Abstract**

The Benes network has been used as a rearrangeable network for over 40 years, yet the uniform $N(2 \log N - 1)$ control complexity of the $N \times N$ Benes is not optimal for many permutations. In this paper, we present a novel $O(\log N)$ depth rearrangeable network called KR-Benes that is *permutation-specific control-optimal*. The KR-Benes routes *every* permutation with the minimal control complexity *specific* to that permutation and its worst-case complexity for arbitrary permutations is bounded by the Benes; thus it replaces the Benes when considering control complexity/latency. We design the KR-Benes by first constructing a restricted $2 \log K + 2$ depth rearrangeable network called $K$-Benes for routing $K$-bounded permutations with control $2N \log K$, $0 \leq K \leq N/4$. We then show that the $N \times N$ Benes network itself (with one additional stage) contains every $K$-Benes network as a subgraph and use this property to construct the KR-Benes network. With regard to the control-optimality of the KR-Benes, we show that any optimal network for rearrangeably routing $K$-bounded permutations must have depth $2 \log K + 2$, and therefore the $K$-Benes (and hence the KR-Benes) is optimal.

## I. INTRODUCTION

The main result of this paper can be summarized as "The Benes network structure is more powerful than imagined". Simply put, in this paper we present the interesting result that by making a few small modifications to the Benes network [1], the resulting rearrangeable network can route most permutations with much lower latency and control overhead. *The worst-case control complexity of the new rearrangeable network for arbitrary permutations is in fact bounded by the Benes.*

Rearrangeable networks such as the Benes contain edge-disjoint paths from inputs to outputs for all $N!$ possible permutations. They find widespread use in shared-memory multiprocessor systems ( [2], [3]), telecommunication networks, time division multiple accessed (TDMA) systems for satellite communication [4] and newer applications such as switching fabrics in internet routers. Many packet switches based on combined input-output queing models [5], require a permutation network in the middle of the switch fabric for routing packets from input queues to output ports during each time slot. High-speed optical switching networks which require crosstalk-free routing also rely on rearrangeable networks [8, 9].

Given their widespread use, improving the control/hardware complexity of rearrangeable networks has tremendous implications. This therefore raises the question: How optimal is the Benes network in this regard? The Benes network is one of the oldest and best-known rearrangeable networks, in use for around 40 years. Waksman showed in [10] that the Benes is the smallest (in terms of depth) $2 \times 2$ switching element based rearrangeable network[1] with a depth[2] of $2 \log N - 1$. Routing in the Benes can be accomplished using the looping algorithm [12], which decomposes a given permutation into two subpermutations that can be routed independently in the subnetworks. The computational time complexity of the looping algorithm (i.e control complexity of the Benes) is $N(2 \log N - 1)$.

There is a vast body of literature examining the rearrangeability of the Benes and other isomorphic MIN based networks and developing new routing algorithms for these networks. Lee [11] proved the rearrangeability of omega-omega$^{-1}$ networks by describing passability conditions for permutations. The rearrangeability of symmetric MINs was studied by Yeh and Feng [14] and Kim et al [13]. These networks have the same hardware and control complexity as the Benes (i.e $O(\log N)$ depth and $O(N \log N)$ control complexity). Networks with much larger depths have also been designed as permutation networks. For example, Koppelman and Oruc [15] describe a self-routing permutation network with $O(\log^3 N)$ depth and $O(N \log^3 N)$ switching elements. This is improved to $O(N \log^2 N)$ switching elements in [16] and $O(\log^2 N)$ depth in [17]. Unlike the Benes which is composed only of simple $2 \times 2$ switching elements these networks have a factor of $O(\log N)$ more depth and switching elements and require sequential binary adders and comparators [15, 16] and hyperconcentrators [17] which can merge large sequences in parallel using high fan-in logic gates. [18] describes self-routing permutation networks based on De-Bruijn Graphs.

Several results on parallel control of the Benes have also been derived. Nassimi and Sahni [7] describe an $O(\log^2 N)$ algorithm using $O(N)$ processors. In [19], Lee and Zheng describe a fast parallel routing algorithm for Benes based networks called Benes group connectors with $K$ active inputs in $O(log^2 K + \log N)$ time on a completely connected computer or the EREW PRAM model with $N$ processors. In [9], Yang et al. prove that an $N \times N$ Benes can route crosstalk free permutations (i.e no switching elements has more than one active input at a given time) in two passes. Their control algorithm requires splitting the original permutation into non-blocking crosstalk free permutations and can potentially be parallelized. However the control complexity of the Benes for each permutation is still $O(N \log N)$.

In an attempt to reduce the control complexity of the Benes, Feng and Seo [20,21] proposed a new routing algorithm called inside-out routing. They adopted a new approach to routing by starting from the center stages and moving outwards. The control complexity of the Benes network using this algorithm is claimed to be $O(N)$. However, Kim, Yoon, and Maeng [22] have refuted the claim made in [21]. They show that the inside-out routing algorithm requires backtracking due to input-output assignment conflicts and its complexity is no longer $O(N)$. They also show that even with backtracking, the modified inside-out routing algorithm may not be able to find conflict-free assignments for all permutations. Thus to the best of our knowledge, there are no existing $O(\log N)$ depth networks using $2 \times 2$ switching elements in the literature (such as the Benes) that improve its control complexity.

Our results in this paper are motivated by the fact that the hardware and control complexity of the Benes is *uniform*

---

[1] Note that other well-known permutation networks such as the Clos and Crossbar [2] have fewer stages than the Benes but they are not based on $2 \times 2$ switching elements. Being non-blocking, these are obviously more powerful than the Benes, but are correspondingly more complex to implement ($O(N^2)$ crosspoints, as opposed to $O(N \log N)$ for the Benes).

[2] All logarithms are to base 2.

for routing *all* permutations. For example, both identity as well as inverse permutations are routed in a similar manner[3] through the Benes, though they are very different in structure. Thus our objective is to design a permutation-specific control-optimal $O(\log N)$ depth rearrangeable network i.e one that optimally routes **every permutation** with minimal control complexity **specific** to that permutation. We show in this paper that such a network can be designed simply by considering bounded permutations. We describe the rearrangeable KR-Benes network whose control complexity is superior to the Benes on average and whose worst case control complexity is bounded by the uniform $N(2 \log N - 1)$ control complexity of the Benes. We design the KR-Benes by first constructing a restricted network called $K$-Benes for rearrangeably routing $K$-bounded permutations, which are defined as permutations satisfying $|\pi(i)-i| \leq k$, for all inputs $i$, with $K$ the smallest power of 2 integer $\geq k$. We show that *any rearrangeable network* for routing $K$-bounded permutations must have depth at least $2 \log K + 2$, $0 \leq K < N/2$ (depth at least $2 \log K + 1$ for $K = N/2, N$). The $K$-Benes satisfies these constraints and is therefore optimal.

We show that every $K$-Benes is contained as a subgraph in an $N \times N$ Benes network with one additional stage. (Not all rearrangeable networks for $K$-bounded permutations have this subgraph property, as we discuss in Section 3). Based on this, we use $K$-Benes networks as building blocks in constructing the permutation-specific control-optimal KR-Benes [3]. In one implementation, the KR-Benes contains $3 \log N - 3$ stages (of which only $\min(2 \log K + 2, 2 \log N - 1)$ stages are used for routing a given permutation). Alternate implementations of the KR-Benes with $2 \log N$ columns of switching elements using multiplexors can also be derived. The control algorithm for the KR-Benes is a simple modification of the Benes looping algorithm and its complexity is $2N \log K$ which is bounded by the $N(2 \log N - 1)$ complexity of the Benes.

The paper is organized as follows: Section 2 contains brief terminology and background on rearrangeable networks. In Section 3 we describe the K-Benes architecture and control algorithm for routing $K$-bounded permutations. Section 4 discuses lower bounds on the depth and control complexity of a $K$-bounded rearrangeable network and the optimality of the $K$-Benes. Section 5 describes the KR-Benes and Section 6 concludes the paper.

## II. BACKGROUND

An $N \times N$ switching network denotes a network for interconnecting $N$ inputs and $N$ outputs. A switching network capable of handling all possible permutations on $N$ is called a permutation network. Permutation requests from traffic inputs arise in many cases, for example in circuit switched networks for telecommunications, combined input-output queueing based packet switches and routers, and cross-talk free optical networks. A permutation network is *rearrangeable* if for any permutation $\pi$, we can construct edge-disjoint paths in the network linking the $i^{th}$ input to the $\pi(i)^{th}$ output for $0 \leq i \leq N-1$. The Benes network is an example of a rearrangeable network.

There are several ways to describe the architecture of the Benes. We describe it in terms of Butterfly and Inverse Butterfly networks. An $N \times N$ Butterfly [2] consists of $\log N$ columns (or stages) of $2 \times 2$ switching elements arranged in the recursive structure shown in Figure 1. An Inverse Butterfly network is the mirror image of a Butterfly.

---

[3]Note that as a stand-alone network, the $K$-Benes is useful when incoming permutations exhibit static boundedness (i.e., the maximum value of $K$ is bounded and does not increase with time). For such scenarios, a $K$-Benes network can be used instead of the standard $N \times N$ Benes with concomitant hardware and control complexity benefits. However, in most realistic cases, the value of $K$ among permutation requests will change dynamically over time. Therefore the real utility of the $K$-Benes lies in its use as a building block for the KR-Benes network.

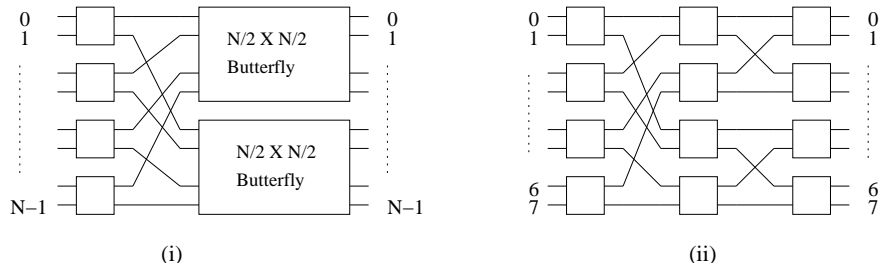

Fig. 1
(I) RECURSIVE STRUCTURE OF THE BUTTERFLY NETWORK (II) AN $8 \times 8$ BUTTERFLY NETWORK.

An $N \times N$ Benes network consists of an $N \times N$ Butterfly followed by an Inverse Butterfly. Thus the Benes contains $2 \log N$ stages of switching elements. However the last stage of the Butterfly network can be merged with the first stage of the Inverse Butterfly to decrease the total number of stages to $2 \log N - 1$. Figure 2 illustrates an $8 \times 8$ Benes. Note that the recursive structure of the Butterfly and Inverse Butterfly networks automatically leads to a recursive decomposition of the Benes network. An $N \times N$ Benes can be viewed as consisting of two outer stages of switching elements connected to top and bottom $\frac{N}{2} \times \frac{N}{2}$ Benes subnetworks. This is very useful in deriving the looping algorithm [2] for routing permutations in the network.

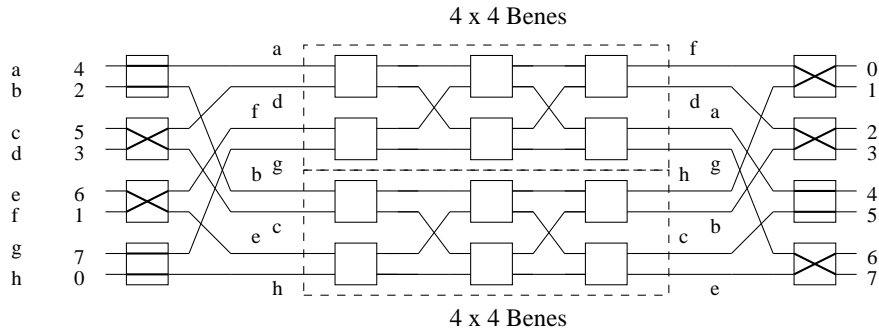

Fig. 2
NETWORK ARCHITECTURE AND CONTROL ALGORITHM FOR AN $8 \times 8$ BENES.

Figure 2 also shows an example of the looping algorithm. It begins by setting a switching element in the outer left stage such that companion inputs are directed to top and bottom subnetworks. The destination switching elements in the outer right stage must automatically be set to receive these inputs from the correct subnetworks. By alternately setting switches in the outer stages, the entire permutation can be divided into two smaller permutations to be realized at each Benes subnetwork. The looping algorithm sequentially examines the inputs in each stage and hence the control complexity of the Benes is $O(N \log N)$. The parallel version of the looping algorithm has complexity $O(\log^2 N)$ using $O(N)$ processors [7].

A $K$-bounded permutation $\pi$ is one which satisfies the condition: $|\pi(i)-i| \leq k$, for all inputs $i$, $0 \leq i \leq N-1$, $k$ is any integer in $[0 \ldots N-1]$ and $K \leq N$ is the smallest power of 2 integer $\geq k$. In general, there are $K!(K+1)^{N-K}$ such permutations, for $K \leq N/2$. (The first $N - K$ inputs have $K + 1$ output choices while the last $K$ together have $K!$ choices).

Given a $K$-bounded permutation, we consider two problems:

- What is the optimal depth rearrangeable network that can route $K$-bounded inputs to outputs (assuming only $2 \times 2$ switching elements)? It seems evident that the optimal network will have depth $O(\log K)$, but the exact constants need to be determined since this will affect control complexity.

- Can this optimal but *restricted* rearrangeable network be designed so as to be an efficient building block for a network that control-optimally routes every permutation? A trivial (but non-efficient) solution is to have $\log N$ parallel copies of the network of problem 1 for $K = 0, 2, 4, 8, \ldots$.

We now describe our rearrangeable network for $K$-bounded permutations, the $N$-input $K$-Benes. In the next section, we show that it is optimal.

Consider any $K$-bounded permutation $\pi$. Divide the set of $N$ input and output lines numbered from $0, 1 \ldots, N-1$, into $N/K$ bands of $K$ contiguous inputs and outputs each. Each such set of input lines is labeled $I_i$, where $I_i = [iK, \ldots, (i+1)K-1]$, $0 \leq i \leq N/K-1$. Similarly, $O_i$ refers to the $i^{th}$ output band. Note that the input and output bands refer to the same set of numbers. Also in our notation, when we refer to input $i$, we are referring to the packet (or connection) on input line $i$, which may be destined to any output. Thus for the given permutation $\pi$, inputs in $I_i$ are destined only to the same or adjacent output bands, i.e., $\pi(j) \in \{O_{i-1}, O_i, O_{i+1}\}, \forall j \in I_i$. For example, in a 4-bounded permutation, (packets on) inputs $0, 1, 2, 3$ in $I_0$ are potentially destined to output bands $O_0$ and $O_1$, i.e., output lines 0 through 7.

For each input band, further define subsets $I_{i,U} \subseteq I_i$ and $I_{i,D} \subseteq I_i$, where $I_{i,U} = \{j : \pi(j) \in O_{i-1}\}$ and $I_{i,D} = \{j : \pi(j) \in O_{i+1}\}$. Also define $I_{i,S} = I_i - (I_{i,U} \bigcup I_{i,D})$. Members of $I_{i,U}$ and $I_{i,D}$ represent *migrating* inputs destined to the upper and lower bands respectively, while members of $I_{i,S}$ represent *stationary* inputs, destined within the same band.

*Observation 1:* $|I_{i,U}| = |I_{i-1,D}|$ and $|I_{i,D}| = |I_{i+1,U}|$, since the number of migrating inputs into, and out of, a band must be equal.

Figure 3 describes a schematic of the proposed $K$-Benes rearrangeable network for a given value of $K$, $0 \leq K \leq N/4$ (the $K$-Benes is identical in performance to the standard Benes for $K > N/4$). The network consists of $2\log K + 2$ columns of $2 \times 2$ switching elements and can be logically divided into three component stages by function. Packets in the network first go through a matching stage implemented via a matching network followed by a band-exchange stage implemented via shuffle-exchange interconnections, and finally a routing stage where packets are routed to their final destinations.

The matching stage of the K-Benes network consists of $N/K$ matching networks $M_i$, stacked over each other. Each matching network is the equivalent of a $K \times K$ Inverse Butterfly network and consists of $\log K$ columns of switching

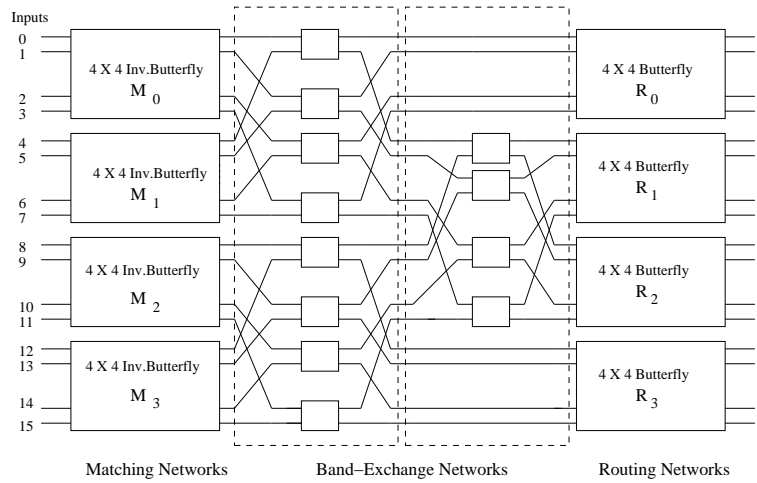

Fig. 3
SCHEMATIC OF AN $N \times N$ K-BENES NETWORK, $N = 16$, $K = 4$.

elements, numbered left to right from 0 through $\log K - 1$.

The band-exchange stage of the $K$-Benes consists of two successive columns of switching elements together labeled as BE$(N, K)$. Each column of a BE$(N, K)$ implements a shuffle-exchange interconnection. The first 'even' band-exchange column implements a shuffle-exchange between the outputs of matching network pairs $(M_0, M_1), (M_2, M_3), \ldots$. The second 'odd' band-exchange column implements shuffle-exchange interconnections between the outputs of matching network pairs $(M_1, M_2), (M_3, M_4), \ldots$ as they come out of the first column. BE(16,2) and BE(16,4) networks are also illustrated in Figure 4.

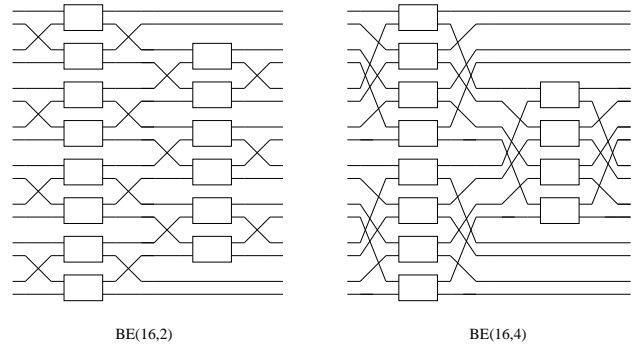

Fig. 4
BAND-EXCHANGE NETWORKS.

Finally, the third logical stage of the $K$-Benes consists of $N/K$ routing networks $R_i$ stacked over each other, where each $R_i$ is a $K \times K$ Butterfly.

The $K$-Benes network has the following property:

*Property 1:* The first $\log K + 1$ together with the last $\log K$ stages of a regular $N \times N$ Benes are isomorphic to a $K$-Benes without the 'odd' band-exchange column in the BE$(N, K)$ network.

An $N \times N$ Benes consists of an $N \times N$ Butterfly followed by an $N \times N$ Inverse Butterfly. The property follows from the fact that the first $\log K + 1$ columns of an $N \times N$ Butterfly can be made equivalent to stacked $K \times K$ inverse butterflies followed by the even band-exchange column of $BE(N,K)$ by relabeling the switching elements in each succeeding column $p$ using an $\frac{N}{2^p}$-way perfect shuffle, $0 \le p \le \log K - 1$. (Please refer to [2] for details). Likewise the last $\log K$ columns of an $N \times N$ Benes can be made equivalent to stacked $K \times K$ Butterflies.

Figure 5 illustrates this isomorphism between the first 3 stages of a Benes and a $K$-Benes followed by even band-exchange network ($K = 4$).

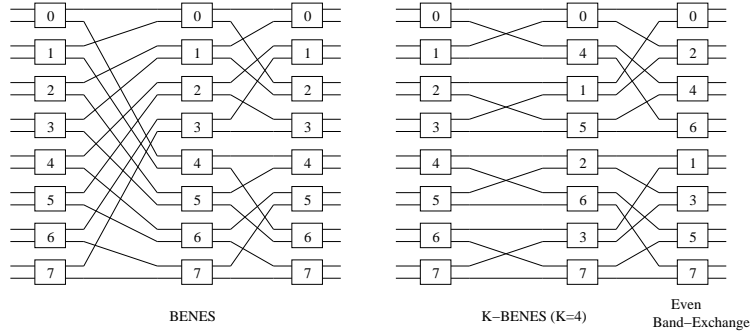

Fig. 5
EXAMPLE ILLUSTRATING ISOMORPHISM OF PROPERTY 1.

Thus an $N \times N$ Benes with one extra stage (the odd band-exchange column of $BE(N,K)$) positioned appropriately, contains a $K$-Benes as an isomorphic subgraph. We use this property later to construct the control-optimal KR-Benes network.

We will now show that the $K$-Benes is rearrangeable for $K$-bounded permutations. The main intuition behind our network design is as follows: Suppose we apply the $N$-input Benes control looping algorithm to set the switches in the Matching and Routing networks of the $K$-Benes. By property 1, the K-Benes is an isomorphic subgraph of the Benes and we can obtain the switch settings of the matching and routing networks via the isomorphism.

Now consider an arbitrary input $\alpha \in I_i$ such that $\pi(\alpha) = \beta$, where $\pi$ is a $K$-bounded permutation. Suppose $\alpha$ appears at output line $j$ of $M_i$ after following the above switch settings at each stage, $0 \le j \le K - 1$. If $\alpha$ is a migrating up (down) input, then following the switch settings from the $j^{th}$ input line of $R_{i-1}$ ($R_{i+1}$) will lead to output $\beta$. Similarly, if $\alpha$ is a non-migrating input, then following the switch settings from the $j^{th}$ input line of $R_i$ will lead to output $\beta$. (We will prove these two statements shortly). In order to prove the rearrangeability of the $K$-Benes, all we have to do is show that a migrating up (down) input $\alpha$ at output line $j$ of $M_i$ is **matched** by a migrating down (up) input at output line $j$ of $M_{i-1}$ ($M_{i+1}$). If this is case, we can set the $j^{th}$ switches of the corresponding $BE(N,K)$ networks to the 'cross' state thereby exchanging each migrating input with its 'matched' counterpart. At this point all migrating inputs are in the correct bands, *at the exact positions they would be in if they had started out in that band* and can be routed to their destinations by simply following the switch settings of the routing network.

Consider an $N$-input $K$-Benes network where the switching elements (SEs) in the $M_i$ and $R_i$ networks are set using the $N$-input looping algorithm. $\alpha \in I_i$, $\beta$, $\pi$ and $j$ are as defined above. Let $b_0 b_1 \ldots b_{\log K - 1}$ denote the routing path of $\alpha$, from left to right through the $\log K$ stages of $M_i$, where $b_m = 0$ ($b_m = 1$) represents whether $\alpha$ is routed to the upper or lower subnetwork in stage $m$ of $M_i$, $0 \leq m \leq \log K - 1$ (i.e via the upper (0) or lower (1) output link of its SE in stage $m$). The following lemma formally proves the above statements.

*Lemma 1:* If $\alpha$ is a migrating up (down) input that appears at output $j$ of $M_i$, then following the switch settings from the $j^{th}$ input line of $R_{i-1}$ ($R_{i+1}$) will lead to output $\beta$. if $\alpha$ is a non-migrating input, then following the switch settings from the $j^{th}$ input line of $R_i$ will lead to output $\beta$.

*Proof:* Since $M_i$ is an inverse butterfly network, we have $b_0 b_1 \ldots b_{\log K - 1} = j$, by definition of the unique path property of the inverse butterfly [6]. Without loss of generality, assume $\alpha$ is a down migrating input. By property 1, the SEs set by the looping algorithm that lead to output $\beta$ are in $R_{i+1}$. Let $\hat{j}$ be the input of $R_i$ that leads to output $\beta$. We now use the fact that the looping algorithm sets switching elements symmetrically, i.e. if an SE for $\alpha$ is set to 0 (1) in stage $m$ of $M_i$, it will be set to 0 (1) in stage $\log K - m$ of $R_{i+1}$. Therefore the established routing path from $\beta$ to $\hat{j}$ is identical to the path from $\alpha$ to $j$. $R_i$ is a butterfly network, however looking at $R_i$ from the output side to the input side, it is an inverse butterfly. Therefore, by the routing property of inverse butterfly networks $\hat{j} = j$. A similar argument holds true when $\alpha$ is an up migrating or non-migrating input. ∎

We formally define the concept of matching inputs below.

*Definition 1:* **Matching Inputs**: Inputs $a \in I_{i,D}$ and $b \in I_{i+1,U}$ are said to be matching if they have the same routing paths (i.e they follow the same sequence of top and bottom SE links) in their respective matching networks. Matching is similarly defined for inputs in $I_{i,U}$ and $I_{i-1,D}$, $0 \leq i \leq (N/K) - 1$.

Thus matching input pairs (if they exist), will be routed to the same output line of their respective matching networks. To prove that every migrating input has a matching pair, we use the following lemma.

*Lemma 2:* Set the switches in the Matching and Routing networks using the looping algorithm. Consider the inputs at an arbitrary stage $l$ of $M_i$, $0 \leq l \leq \log K - 1$. Let $I_{i,U}^T$ and $I_{i,U}^B$ ($I_{i,D}^T$ and $I_{i,D}^B$, resp.) represent inputs in $I_{i,U}$ ($I_{i,D}$ resp.) routed via upper (lower, resp.) output links of the SE in stage $l$. Then we have,

$$|I_{i,U}^T| = |I_{i-1,D}^T| \qquad (1)$$
$$|I_{i,U}^B| = |I_{i-1,D}^B| \qquad (2)$$
$$|I_{i,D}^T| = |I_{i+1,U}^T| \qquad (3)$$
$$|I_{i,D}^B| = |I_{i+1,U}^B| \qquad (4)$$

*Proof:* Please see Appendix. ∎

*Theorem 1:* For every migrating input in $I_{i,D}$ and $I_{i,U}$, $0 \leq i \leq \frac{N}{K} - 1$, there exists a matching input in $I_{i+1,U}$ and $I_{i-1,D}$ respectively.

*Proof:* By the matching lemma, at each stage of the K-Benes, the same number of migrating inputs (of each category: up or down) in $I_i$ have been routed via an upper/lower SE link to the next stage as migrating inputs (of the same category) in $I_{i-1}$ and $I_{i+1}$. Therefore for each up/down migrating input at the outputs of $M_i$, there exists a down/up migrating input in $I_{i-1}/I_{i+1}$ which has been routed through the same sequence of upper and lower SE links and therefore appears at the same output position in $M_{i-1}/M_{i+1}$, i.e these are matching inputs. ∎

This leads us to the main result in this section.

*Theorem 2:* The proposed K-Benes network is rearrangeable.

We summarize the steps in the $K$-Benes permutation routing algorithm below.

- Route the inputs in each matching network of the $K$-Benes using the control settings obtained by executing the $N$-input looping algorithm.
- Exchange each migrating input in $I_{i,U}$ and $I_{i,D}$ with a matching input in $I_{i-1,D}$ and $I_{i+1,U}$, respectively.
- Route inputs over each routing network using the control settings obtained previously by the looping algorithm.

Both the K-Benes network architecture and the control algorithm are straightforward. The only 'hard' part of the process lies in recognizing that $K$-bounded permutations can be routed by dividing the inputs into bands and noting that migrating inputs can be matched in adjacent bands when using the Benes control algorithm.

Figure 6 illustrates permutation routing within a $16 \times 16$ K-Benes for $K = 4$. The switch settings are obtained by considering the first and last two stages of the isomorphic Benes network. In this example, $M_0$ consists of two migrating inputs (packets 4 and 5 at input lines 0 and 2, respectively). Note that packet 5 appears at ouput line 4 of $M_0$ as does packet 1 of $M_1$, as claimed by lemma 2. Routing from input lines 3 of $R_0$ and $R_1$ lead to outputs 1 and 5, as claimed in lemma 1. Some of the matching input lines and routing paths are drawn boldfaced for clarity.

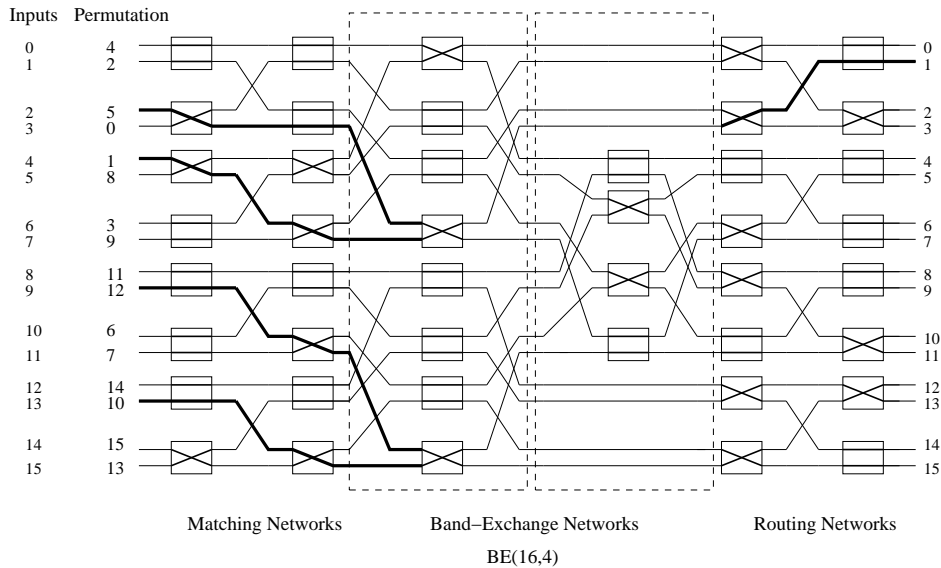

Fig. 6

K-BENES PERMUTATION ROUTING ($K = 4$, $N = 16$).

## IV. OPTIMALITY OF THE K-BENES NETWORK, $0 \leq K \leq N/4$

The depth of a switching network is the number of columns or stages of switching elements. We will show in this section that any rearrangeable network for $K$-bounded permutations must have a depth at least $2 \log K + 2$ and therefore the $K$-Benes is optimal (For simplicity, we will assume $K \leq N/4$ in our discussions since the $K$-Benes reduces to the Benes for $K = N/2$ and $K = N$). Also note that we only consider rearrangeable networks based on simple $2 \times 2$ switching elements, similar to the Benes. We do not consider networks based on comparators or higher degree multiplexors and demultiplexors.

Any rearrangeable network must be able to route permutations containing the following $K$-bounded subpermutations:

1) $\pi_1 : \pi_1(j) = j+K, \forall j \in I_i$, where $i$ is even. Since $\pi$ is $K$-bounded this implies that all inputs in adjacent bands (starting from the $I_0, I_1$ pair) are destined to each other.
2) $\pi_2 : \pi_2(j) = j-K, \forall j \in I_i$, where $i$ is even, $i \neq \{0, (N/K)-1\}$.
3) $\pi_3 : [iK, \ldots, (i+1)K-1] \to [iK, \ldots, (i+1)K-1]$ for a given input band $I_i$. Thus $\pi_3$ is an arbitrary permutation within a given band.

Define an even band-exchange network as a set of switches for interchanging inputs $j$ and $j + K$ for all such inputs in even band pairs $(I_0, I_1), (I_2, I_3), \ldots$. Likewise define an odd band-exchange network for inputs in band pairs $(I_1, I_2), (I_3, I_4), \ldots$. Henceforth we will use the term 'BE switch' to refer to a single switching element that implements an exchange between adjacent bands. Note that the above definition does not restrict all BE switches to be together in the same column (they could be dispersed through the rest of the network). However, when all the BE switches are together, these networks are identical to the columns of the BE$(N, K)$ network defined in the last section. An even band-exchange network can implement $\pi_1$ while an odd band-exchange network can implement $\pi_2$. $\pi_3$ can be implemented optimally by a $K \times K$ Benes network. Moreover, none of these three subnetworks can implement any of the other two subpermutations.

*Lemma 3:* Any rearrangeable network for $K$-bounded permutations must contain an even band-exchange stage, an odd band-exchange stage and a $K \times K$ Benes network as subgraphs.

*Proof:* Any $K$-bounded permutation, $0 \leq K \leq \frac{N}{4}$, may contain one or more of the three permutations above as subpermutations. Therefore any rearrangeable network must contain an even band-exchange stage, an odd band-exchange stage and a $K \times K$ Benes network as subgraphs. ∎

*Lemma 4:* The optimal rearrangeable network for $K$-bounded permutations has exactly two band-exchange networks as a subgraph.

*Proof:* By symmetry, the optimal network must have the same number of even and odd band-exchange stages. From lemma 3, a total of two band-exchange stages is necessary. Four such stages would increase the minimum depth of the optimal network to $2 \log K + 3$ stages which is more than the K-Benes, a contradiction. Therefore the optimal rearrangeable network contains exactly two band-exchange networks as a subgraph. ∎

Lemma 4 implies that the position of the BE switches within the optimal rearrangeable network is critical, since there can be **only one** band-exchange network of either type. Again note that the lemma does not restrict the optimal network to having the BE switches for all $N/K$ bands together at the same depth (i.e equivalent to a BE$(N,K)$). It merely states that there can be no more than two total band-exchange connections for each band (equivalently, exactly two BE switches at a given line number $j$, one for exchanging with either adjacent band). However, as we show below, BE switches (for any of the $N/K$ bands) cannot *all* appear 'too early' in terms of their depth in the optimal network.

*Lemma 5:* The optimal rearrangeable network contains at least one BE switch after $\log K$ stages, i.e stage $\log K +1$

onwards.

*Proof:* We prove this statement by contradiction. Assume the last BE switch for inputs in bands $I_i$ and $I_{i+1}$ occurs at depth $\log K$ (i.e after $\log K - 1$ stages) or earlier. Note that we are considering rearrangeable networks consisting only of $2 \times 2$ switching elements. Any multistage interconnection network (MIN) composed of $2 \times 2$ switching elements can be viewed as a union of complete binary trees, with switching elements as nodes. Consider a down migrating input at input line $a \in I_i$ (i.e $\pi(a) \in O_{i+1}$). From the standard properties of binary trees, input line $a$ in such a network can reach at most $K/2$ output lines over $\log K - 1$ stages. Let $OX_i$ be this reachable set for input line $a$. Without loss of generality (WLOG), consider the identical set of lines $OX_{i+1}$ in the next band $I_{i+1}$. Using the BE switch under consideration, the input on line $a$ can only be exchanged with one of the $K/2$ inputs in $I_{i+1}$ that can reach set $OX_{i+1}$. If none of these inputs are up migrating inputs, then this particular exchanged input can never be sent back into its destination band $O_{i+1}$ as there are no BE switches after stage $\log K$. Hence this input line will never be routed to its destination which contradicts the claim that the network is rearrangeable. Therefore the assumption must be incorrect, which proves the statement of the lemma. ∎

*Lemma 6:* The optimal rearrangeable network contains at least one BE switch after $\log K + 1$ stages, i.e stage $\log K + 2$ onwards.

*Proof:* The discussion in lemma 5 referred to bands $I_i$ and $I_{i+1}$. It independently applies to bands $I_i$ and $I_{i-1}$ whose last BE switch(es) cannot be in the same stage as the BE switch(es) for bands $I_i$ and $I_{i+1}$. They must be at least one stage later and hence the lemma. ∎

WLOG using lemma 6, let there be a BE switch in stage $\log K + 2$ of the optimal rearrangeable network.

*Lemma 7:* The optimal rearrangeable network contains at least $\log K$ stages after stage $\log K + 2$.

*Proof:* Consider bands $I_i$ and $I_{i+1}$. Figure 7 illustrates a possible routing situation after $\log K - 1$ stages in the optimal rearrangeable network. In the figure, $a$ is the solitary down migrating input in $I_i$ with $b$ the solitary up migrating input in $I_{i+1}$. WLOG, the set of output lines reachable from input line $a$ are the first $K/2$ lines of $O_i$ (they could be any set of $K/2$ lines but we can assume they are the first $K/2$ lines for simplicity, without affecting the proof). Denote by $X_a$, the set of input lines (including $a$) which can reach this output set. Note that $|X_a| = K/2$ because of standard binary tree based MIN properties. The remaining $Y_a$ input lines of $I_i$ reach the last $K/2$ lines of $O_i$, after $\log K - 1$ stages, $|Y_a| = K/2$. Note that these $Y_a$ lines do not contain any migrating inputs. Similarly, assume input line $b$ is positioned in $I_{i+1}$ such that only the last $K/2$ lines of $O_{i+1}$ are reachable from line $b$ and define $X_b$ and $Y_b$ similarly, where none of the $Y_b$ lines contain any migrating inputs.

Assume input line $a$ (input $b$) can be exchanged with inputs from $Y_b$ ($Y_a$, resp.) in this stage *if necessary*, i.e there are BE switches present in this $\log K^{th}$ stage. The optimal rearrangeable network algorithm can position $a$ anywhere within the first $K/2$ lines and exchange it with the most suitable input line from $Y_b$. Let this line be $\alpha$. Likewise, let $\beta \in Y_a$ be the input that can be exchanged with $b$ in this stage. We will show later that stage $\log K$ is the earliest stage in which $a$ and $\alpha$ ($b$ and $\beta$, resp.) can be exchanged.

Now note that if $a$ and $\alpha$ are exchanged in stage $\log K$, the earliest instance $\alpha$ can be exchanged again is stage $\log K + 3$. Let $a$ and $\alpha$ be exchanged by BE switch $j$. First, we can assume that the SE in position $j$ of stage $\log K + 1$

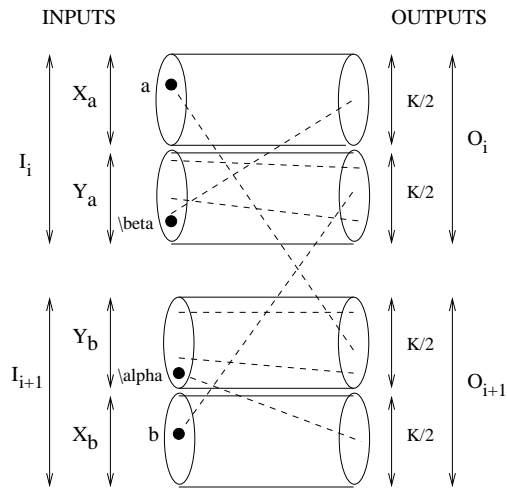

Fig. 7

ILLUSTRATION FOR LEMMA 7. DASHED LINES IN THE FIGURE INDICATE DESTINATION LOCATIONS OF INPUTS. BOLD LINES INDICATE THE REACHABILITY OF OUTPUT LINES FROM THE GIVEN INPUT LINES.

is a band-exchange switch for bands $I_i$ and $I_{i-1}$ (by symmetry). Next, by lemma 4, the optimal network cannot contain another BE switch in position $j$. Therefore $\alpha$ must move to some other position before being input to another BE switch, which requires at least one stage. Therefore $\alpha$ cannot be exchanged with any other input from $I_{i+1}$ before stage $\log K + 3$.

Now we have four possibilities:

1) Only $a$ and $\alpha$ are exchanged in stage $\log K$.
2) Only $b$ and $\beta$ are exchanged in stage $\log K$.
3) Both pairs are exchanged in stage $\log K$.
4) None of the pairs are exchanged in stage $\log K$.

If case 4 is true, assume WLOG that $a$ and $b$ are exchanged with each other. This can be done only in stage $\log K + 1$ or stage $\log K + 2$ (for the symmetric case of $a$ and $b$ in bands $I_i$ and $I_{i-1}$). This is because $a$ and $b$ are present in complementary sets of output lines at the input to stage $\log K$ (as shown in Figure 7) and one more stage is required for them to match up (two more stages for the symmetric case). Once $a$ and $b$ are exchanged (effectively in stage $\log K + 2$), they are in the same relative position within their bands. In $\log K - 1$ stages after this point, $a$ and $b$ can reach a maximum of $K/2$ output lines. If $a$ is destined to one set of $K/2$ output lines and $b$ is destined to the complementary set of $K/2$ lines, then they cannot **both** be routed to their destinations using only $\log K - 1$ stages. In this case, either $\log K$ more stages are necessary (which proves the lemma) **or** the optimal rearrangeable network (algorithm) can choose to exchange $a$ with $\alpha$ first (case 1), so that $a$ is now closer to its destination. However, now $\alpha$ is in the wrong band and must be exchanged with $b$ or with $\beta$ (case 3) in stage $\log K + 3$ as shown earlier. At least $\log K - 1$ more stages are required after this point to route $\alpha$ to its destination, thereby proving the lemma. An analgous analysis holds for case 2.

We now show that the stage $\log K$ is the earliest at which $a$ and $\alpha$ can be exchanged (followed by $\alpha$ and $b$ later as described above). Suppose $a$ is destined to an output among the top $K/2$ of output lines in $O_{i+1}$. Let all the inputs in $Y_b$, except one, be destined to the remaininder of these $K/2$ output lines. Label this exceptional input $\alpha$. $\alpha$ is destined to the bottom half of the the output lines in $O_{i+1}$. $b$ is also destined to the bottom half of output lines in $O_i$ (note that this is the necessary condition for case 1 to occur), and therefore if $\alpha$ and $b$ are exchanged they can reach their destinations in $\log K - 1$ stages. If any other input line from $Y_b$ is exchanged with $a$, then that input and $b$ cannot be routed to their output lines in $\log K - 1$ stages which will increase the total number of stages in the optimal network to $2 \log K + 3$, a contradiction. Now $\alpha$ and $a$ must both arrive at position $j$ in their respective bands to be exchanged. By the property of binary tree based MINs it takes exactly $\log K - 1$ stages to achieve this. ∎

*Theorem 3:* The $K$-Benes is an optimal rearrangeable network for $K$-bounded permutations.

*Proof:* The result follows from lemmas 1–lemma 7. ∎

## V. THE KR-BENES: A CONTROL-OPTIMAL REARRANGEABLE NETWORK

We now describe how the $K$-Benes can be used as a building block to construct the rearrangeable KR-Benes network that is control-optimal for all input permutations. The KR-Benes is divided into regular Benes stages and Band-Exchange network stages.

Consider the $N \times N$ Benes network with $2 \log N - 1$ stages numbered from left to right from 1 onwards. The KR-Benes is constructed using the following steps:

1) Insert Band-Exchange networks BE$(N, 2^i)$ consisting of even and odd band-exchange columns immediately at the outputs of each stage $i$, $1 \leq i \leq \log N - 2$ of the Benes (after renumbering them as per the isomorphism with the inverse Butterfly illustrated in Figure 5).

2) We insert two kinds of bypass edges in the network that will allow an input to bypass some stages in the underlying network.

    a) First attach bypass edges leading from the outputs of the BE$(N, 2^i)$ network directly to the inputs of corresponding switching elements in stage $2 \log N - i$ of the Benes (also renumbered as per the isomorphism with the inverse Butterfly).

    b) Next attach bypass edges at the output of each Benes stage to the input of the next stage that allow each Band-Exchange network itself to be bypassed.

In a sense, the KR-Benes network consists of multiple planes: The front plane is the $N \times N$ Benes network while the BE$(N, 2^i)$ networks are in a (bypassable) parallel backplane between each of the first $\log N - 2$ stages of the Benes. Figure 8 describes an $8 \times 8$ KR-Benes. Note that there is only one Band-Exchange network in this case (a BE(8,2) after stage 1).

*Theorem 4:* An $N \times N$ KR-Benes contains every $K$-Benes network as a subgraph, $0 \leq K \leq N/4$.

*Proof:* By observation 1 in section 3, we know that the first and last $\log K$ stages of an $N \times N$ Benes together are equivalent to a $K$-Benes without its middle BE$(N, K)$ subnetwork. Therefore, the first $\log K$ stages of the KR-Benes (with the first $\log K - 1$ BE() networks bypassed) followed by the BE$(N, K)$ network and then the last $\log K$ stages of the KR-Benes (reached via the bypass lines at the outputs of the BE$(N, K)$ network) clearly form a $K$-Benes network. Thus the subgraph property is satisfied. ∎

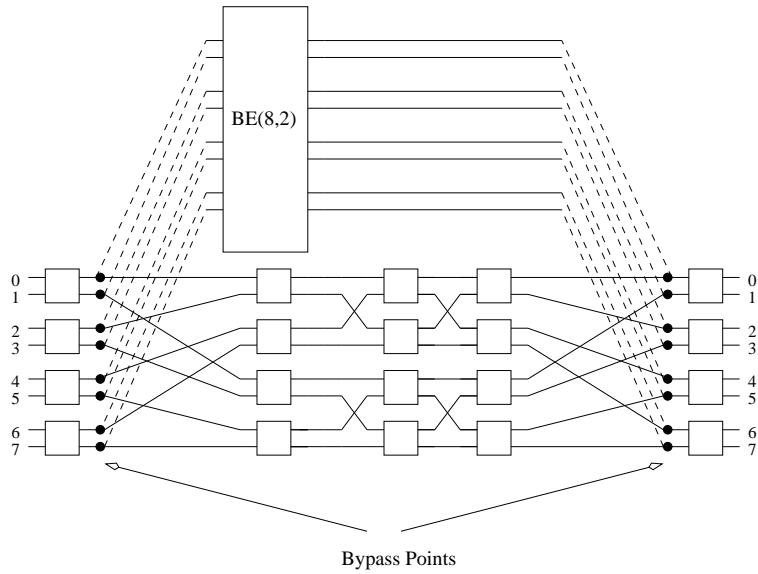

Fig. 8
$8 \times 8$ KR-BENES.

*Corollary 1:* The control complexity of the $N \times N$ KR-Benes is $\min(N(2 \log K), N(2 \log N - 1))$ for any arbitrary permutation $\pi$ on $N$ inputs, where $K = \max(|\pi(i) - i|)$, $0 \leq i \leq N-1$ ($K$ is rounded up to the nearest power of two).

Note that the KR-Benes as described above contains $4 \log N - 5$ stages ($2 \log N - 1$ Benes stages and $2(\log N - 2)$ Band-Exchange columns). A simple optimization can reduce the total number of stages in the KR-Benes to $3 \log N - 3$ ($2 \log N - 1$ Benes stages and $\log N - 2$ Band-Exchange columns). Consider the $N \times N$ Benes network constructed as a concatenation of $N \times N$ inverse butterfly and butterfly networks. Insert the 'odd' band-exchange column from $BE(N, 2^i)$ immediately at the output of stage $i + 1$ of this Benes, $1 \leq i \leq \log N - 2$. Using the fact that stage $i$ of this Benes is isomorphic to the 'even' band-exchange column of $BE(N, 2^i)$ (observation 1) and inserting appropriate bypass lines, we note that this implementation of the KR-Benes also contains every $K$-Benes as a subgraph and hence is control-optimal.

Further note that only one of the Band-Exchange networks will ever be used for routing any permutation. Hence an alternate implementation of the KR-Benes with $2 \log N$ columns of switching elements can be derived by using a single odd Band-Exchange network (simulating different values of $K$) in the backplane to which outputs from various stages of the frontplane $N \times N$ Benes are multiplexed. However this scheme will require the use of several multiplexors at the inputs and outputs of the Band-Exchange column.

Every $K$-Benes network is contained in a KR-Benes and the $K$-Benes control algorithm is the looping algorithm for the isomorphic Benes in its first and last $\log K$ stages. Therefore for an arbitrary input permutation, we only need to simulate $K$-Benes routing in the appropriate subgraph of the KR-Benes. The control-algorithm for the KR-Benes is as follows:

- Run the Benes looping algorithm for the first (and last) stage of the KR-Benes. In the process, determine the

value of $K$.

- Use the above value of $K$ to select bypass lines for the first $\log K - 1$ Band-Exchange networks Note that the bypass lines can be selected in a self-routing manner after $K$ has been determined while controlling the first stage. Also mark those packets which are migrating inputs.
- Exchange migrating inputs in the BE$(N, K)$ subnetwork. Since the inputs to be exchanged are marked, the BE(.) networks are self-routing.
- Select bypass lines at the output of BE$(N, K)$ to reach the $(2 \log N - K)^{th}$ Benes stage of the KR-Benes and route the packets to their destinations.

For example, a K-Benes with $K = 4$ can be simulated by bypassing the first BE$(N, 2)$ network, passing through the BE$(N, 4)$ and then reaching the secondlast stage of the KR-Benes directly via bypass lines. Thus the control complexity of the KR-Benes is the same as that of of the corresponding K-Benes plus the constant overhead required to set the bypass links and BE(2) switches (which are both self-routing).

The control complexity of the KR-Benes is superior to the Benes on average and its worst case is bounded by the Benes whose control complexity is $N(2 \log N - 1)$. We can evaluate the performance of the KR-Benes using the following metric: There are $K!(K+1)^{N-K}$ $K$-bounded permutations. The KR-Benes control complexity for a given $K$-bounded permutation is $2N \log K$, $0 \leq K \leq N/4$ and $N(2 \log N - 1)$, otherwise. Thus the KR-Benes has superior control complexity as compared to the Benes for $P_N = (\frac{N}{4})!(\frac{N}{4} + 1)^{\frac{3N}{4}}$ permutations.

In an environment where all incoming permutations are equally likely, the average control complexity of the KR-Benes can be computed as follows: Let $P_K$ represent the number of $K$-bounded permutations that are not $K/2$-bounded. We have $P_K = K!(K+1)^{N-K} - (\frac{K}{2})!(\frac{K+2}{2})^{N-\frac{K}{2}}$. The the average control complexity of the KR-Benes is:

$$1 + (2N \sum_{K=2}^{N/4} P_K \cdot \log K) + N(2 \log N - 1)(N! - P_N)$$

In many practical cases, the distribution of incoming $K$-bounded permutations to the KR-Benes is likely to be highly non-uniform and biased towards smaller $K$ (due to locality properties), which enhances the control gain of the KR-Benes. Solutions for P2P file sharing, P2P overlay networks and P2P distributed hash tables such as CAN, Viceroy, Kademlia etc. emphasize locality of access to increase throughput. Permutation networks serving routing needs for such networks are likely to receive a large number of bounded permutations. Among other examples, a permutation network serving as an optical switch [9] in a high-speed LAN environment might typically get most requests to and from addresses within the LAN with a few periodically outside the local area. Thus in these cases smaller bounded permutations are more likely as inputs, which can be routed more efficiently by the KR-Benes.

## VI. Conclusion

This paper demonstrates the power and elegance of the Benes for rearrangeably routing permutations. By making small modifications to the Benes in order to exploit the locality property of permutations (also called $K$-boundedness), we can design a new network with $3 \log N - 3$ (alternately $2 \log N$) stages, that is control-optimal for arbitrary permutations. Its control complexity is superior to the Benes in many cases ($\approx (\frac{N}{4})!(\frac{N}{4}+1)^{\frac{3N}{4}}$ permutations) and bounded by the Benes in the worst case. The K-Benes can be used for static traffic scenarios while the KR-Benes network can be used for optimally routing $K$-bounded permutations, where the value of $K$ changes dynamically.

## References

[1] V. E. Benes, "Optimal Rearrangeable Multistage Connecting Networks," Bell System Technical Journal, 43 (1964), pp. 1641-1656.
[2] F. T. Leighton. Introduction to Parallel Algorithms and Architectures: Arrays, Trees, Hypercubes. Morgan Kaufmann Publishers, San Mateo, CA, 1992.
[3] David Nassimi, Sartaj Sahni, "A Self-Routing Benes Network and Parallel Permutation Algorithms," *IEEE Transactions on Computers*, 30(5): 332-340 (1981)
[4] S. Keshav. An Engineering Approach to Compter Networking. Addison-Wesley Publishers, 1997.
[5] B. Prabhakar and N. McKeown, "On the Speedup Required for Combined Input and Output Queued Switching," *Automatica*, 35(12):1909-1920, 1999.
[6] J. Duato and L. Ni, "Interconnection Networks", IEEE Computer Society Press, Los Alamitos, CA., 1999.
[7] David Nassimi and Sartaj Sahni, "Parallel Algorithms to Set Up the Benes Permutation Network," *IEEE Transactions on Computers*, 31(2): 148-154 (1982).
[8] A. Pattavina and G. Maier, "Photonic Rearrangeable Networks with Zero Switching-Element Crosstalk," *IEEE INFOCOM 1999*, pp. 337-344, April 1999.
[9] Y. Yang, J. Wang and Y. Pan, "Permutation Capability of Optical Multistage Interconnection Networks," *J. of Parallel and Dist. Computing*, Vol. 60, No. 1, pp. 72-91, Jan. 2000.
[10] A.Waksman, "A Permutation Network," *Journal of the ACM*, Vol.15, No.1, pp.159-163, Jan. 1968.
[11] K. Y. Lee, "On the Rearrangeability of $(2 \log N - 1)$ Stage Permutation Networks, *IEEE Trans. Computers*, 34(5): 412-425 (1985).
[12] D.C. Opferman and N.T. Tsao-Wu, "On a Class of Rearrangeable Switching Networks, Part I: Control Algorithm", *Bell System Technical J.*, Vol.50, pp.1,579-1,600, 1971.
[13] M. K. Kim, H. Yoon and S. R. Maeng, "Bit-Permute Multistage Interconnection Networks," *Microprocessing and Microprogramming*, Vol. 41, pp. 449–468, 1995.
[14] Y-M. Yeh, T-Y. Feng, "On a Class of Rearrangeable Networks," *IEEE Trans. Computers*, 41(11): 1361-1379 (1992).
[15] D. M. Koppelman, A. Y. Oruc, "A Self-Routing Permutation Network," *J. Parallel Distrib. Comput*, Vol. 10, No. 2, pp. 140-151, (1990).
[16] C. Jan and Y. Oruc, "Fast Self-Routing Permutation Switching on an Asymptotically Minimum Cost Network," *IEEE Trans. on Comput.*, pp. 1369-1379, Dec. 1993.
[17] H. Cam and J. A. B. Fortes, "A Fast VLSI-Efficient Self-Routing Permutation Network," *IEEE Trans. Computers*, 44(3): 448-453 (1995).
[18] A. Samsudin and K. Y. Lee, " nD-dBPN: New Self-Routing Permutation Networks Based On the de Bruijn Digraphs," in *Intnl. Conf. on Parallel Processing*, pp. 604-611, 1998.
[19] E. Lu and S. Q. Zheng, "A Fast Parallel Routing Algorithm for Benes Group Switches," *Proc. of the 14th IASTED Intl. Conf. on Parallel and Distributed Computing and Systems*, pp. 67-72, Cambridge, MA, Nov. 2002.
[20] S-W. Seo, T-Y. Feng and H-I.Lee, "Permutation Realizability and Fault Tolerance Property of the Inside-Out Routing Algorithm," *IEEE Transactions on Parallel and Distributed Systems*, Vol. 10, No. 9: pp. 946-957, 1999.
[21] T-Y. Feng and S-W. Seo, "A New Routing Algorithm for a Class of Rearrangeable Networks," *IEEE Transactions on Computers*, Vol. 43, No. 11: pp. 1270–1280, 1994.
[22] M. K. Kim, H. Yoon and S. R. Maeng, "On the Correctness of Inside-Out Routing Algorithm," *IEEE Transactions on Computers*, Vol. 46, No. 7, pp. 820-823, 1997.

# Appendix

APPENDIX

## A. Preliminary definitions for proving lemma 2

**Permutation Graph**: Let $\pi : [0,..,N-1] \to [0,..,N-1]$ denote a permutation on $N$ inputs. We define the permutation graph $\Pi$ as a bipartite graph on vertices $V_1 = [0,\ldots,N-1]$ and $V_2 = [0,\ldots,N-1]$ with an edge between $i \in V_1$ and $j \in V_2$ if $j = \pi(i)$. For the $K$-Benes network, we group pairs of adjacent vertices in $V_1$ and $V_2$ to logically represent switching elements (SEs). Let $S_i$ denote the SE of input $i$. $\bar{i}$ denotes its companion input to the SE, where $\bar{i} = i-1$ if $i$ is odd, and $i+1$ otherwise. Thus, there is an edge in $\Pi$ from each switching element $S_i$ to switching elements $S_{\pi(i)}$ and $S_{\pi(\bar{i})}$. A path in the permutation graph alternately traverses switching elements on the input and output side. Fig. 9 illustrates the permutation graph on 8 inputs for the 4-bounded permutation $\begin{pmatrix} 0 & 1 & 2 & 3 & 4 & 5 & 6 & 7 \\ 4 & 5 & 0 & 6 & 1 & 2 & 7 & 3 \end{pmatrix}$. Note that the looping algorithm can be viewed as setting SEs to alternate up and down subnetworks while following a path in $\Pi$.

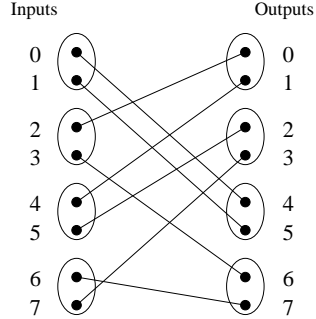

Fig. 9

PERMUTATION GRAPH FOR THE GIVEN PERMUTATION.

Consider a $K$-Benes network with $K = \frac{N}{2}$ and a permutation graph $\Pi$. Note thate there are only two input bands $I_0$ and $I_1$.

**Compatibility Graph**: We construct a compatibility graph $G = (V, E)$ based on the permutation graph $\Pi$ as follows: Each edge in $\Pi$ between a migrating input in $I_0$ or $I_1$ to an output in $O_1$ or $O_0$ is represented as a vertex in $V$. $V$ consists of two subsets $V_1 = \{(a, \pi(a)) | a \in I_{0,D}\}$ and $V_2 = \{(b, \pi(b)) | b \in I_{1,U}\}$.

There are two kinds of edges between vertices in $V$. There exists a *cross* edge labeled $i$ between vertices $u_1 = (a, \pi(a)) \in V_1$ and $u_2 = (b, \pi(b)) \in V_2$ if there is a path of odd length in $\Pi$ between $S_a$ and $S_{\pi(b)}$ or between $S_b$ and $S_{\pi(a)}$ consisting only of switching elements in $I_i$ and $O_i$, $i \in \{0, 1\}$. Similarly, there exists a *straight* edge labeled $i$ between these vertices if $a$ and $b$ are in the same input band and there is a path of even length in $\Pi$ between $S_a$ and $S_b$ or $S_{\pi(a)}$ and $S_{\pi(b)}$ consisting only of switching elements in $I_i$ and $O_i$, $i \in \{0, 1\}$. Figure 10 illustrates the compatibility graph corresponding to the given permutation on 8 inputs. The cross edge between vertices (3,6) and (4,1) is labeled 0 since it arises due to a path of length one in $I_0$. The straight edge between (0,4) and (1,5) is labeled 01 since it arises due to a path of length zero in both $I_0$ and $I_1$.

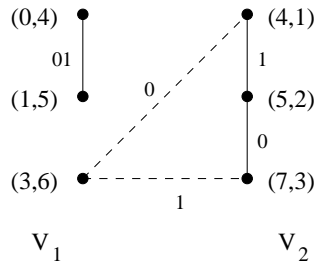

Fig. 10

COMPTABILITY GRAPH CORRESPONDING TO $\Pi$. CROSS EDGES ARE SHOWN DASHED.

A compatibility graph $G$ has the following properties:

*Property 2:* If vertices $(a, \pi(a))$ and $(b, \pi(b))$ are connected by a cross edge in $G$, then $a$ and $b$ are both directed to top (bottom) subnetworks in the next stage via upper (lower) links of their SEs.

*Proof:* Vertices sharing a cross edge are connected by a path of odd length in $\Pi$. The looping algorithm for $\Pi$ sets SEs on a path to alternate top and bottom subnetworks. ∎

Similarly, we have

*Property 3:* If vertices $(a, \pi(a))$ and $(b, \pi(b))$ are connected by a straight edge in $G$, then $a$ and $b$ are directed to opposite subnetworks in the next stage via opposite links of their SEs.

Finally,

*Property 4:* The number of vertex pairs in $V_1$ and $V_2$ connected by straight edges with same label (i.e 0 or 1) are identical.

*Proof:* If a vertex in $V_1$ ($V_2$) is not connected to a vertex in $V_2$ ($V_1$) by a cross edge labeled 0 (1, resp.), then it must be connected to another vertex in $V_1$ ($V_2$) by a straight edge labeled 0 (1, resp.). Hence there must be an identical number of such pairs in $V_1$ and $V_2$. ∎

## B. Proof of lemma 2

Consider a $P \times P$ Benes network implementing a $K$-bounded permutation $\pi$. Set the switches in the first stage of the Benes using the looping algorithm which alternately routes members of each input and output pair to top and bottom subnetworks in the next stage. By the isomorphism of the Matching networks of a $K$-Benes with the first $\log K$ stages of the Benes, proving the lemma is equivalent to proving the stated lemma conditions for the first stage of the above Benes network.

We prove the matching lemma by induction on $P$. For the base case, consider $P = 2K$, i.e there are only two bands $I_0$ and $I_1$. In this case, we need to show that $|I_{1,U}^T| = |I_{0,D}^T|$ and $|I_{1,U}^B| = |I_{0,D}^B|$. To prove this, consider only edges labeled 0 in the compatibility graph for $\pi$. By property 2, all vertices connected by cross edges represent (matching) inputs directed to the same top or bottom subnetwork. By properties 3 and 4, all inputs in $V_1$ connected by straight edges labeled 0 have a (matching) input in $V_2$ that is directed to the same (top or bottom) subnetwork. Therefore the lemma conditions in the base case of the induction hypothesis hold.

Now consider any $K$-bounded permutation on $P$ inputs and assume the inductive hypothesis holds for fewer than $P$ inputs. We will need to find an exact matching between inputs in $I_{0,D}$ and $I_{1,U}$. Assume for the moment that such a matching exists and let $y_i \in I_{0,D}$ be the matching input for each $x_i \in I_{1,U}$, $1 \leq i \leq |I_{1,U}|$. Consider the permutation $\widehat{\pi}$ on $P - K$ inputs defined as follows:

$$\begin{aligned} \widehat{\pi}(x_i) &= \pi(y_i) & 1 \leq i \leq |I_{1,U}| \\ \widehat{\pi}(a) &= \pi(a) & \text{for all other inputs} \quad a. \end{aligned}$$

$\widehat{\pi}$ is a $K$-bounded permutation and hence by the inductive hypothesis, the matching lemma is satisfied for $P-K$ inputs. Inputs $x_i \in I_{1,U}$ are thus assigned top/bottom subnetworks in a manner consistent with all other inputs. Matching inputs $y_i \in I_{0,D}$ can now be assigned the same subnetworks, followed by the remaining inputs in $I_0$, without conflict. Thus the conditions of the matching lemma are satisfied for any set of $P$ inputs and permutation $\pi$.

To complete the proof, we need to find an exact matching between inputs in $I_{0,D}$ and $I_{1,U}$. First construct the compatibility graph $G$ on down migrating inputs in $I_0$ and up migrating inputs in $I_1$ *using only edges labeled 0*. Inputs in $I_0$ represented in $V_1$ with a cross edge to $V_2$ have found a matching input in $I_1$ that is directed to the same subnetwork in the next stage. The remaining vertices in $V_1$ and $V_2$ are connected in pairs by straight edges. We need to find an exact matching for the migrating inputs in $I_0$ and $I_1$ that are represented in these remaining vertices.

Consider one such pair $(x_1, \pi(x_1))$ and $(x_2, \pi(x_2))$ connected by a straight edge in $V_2$ along with $(y_1, \pi(y_1))$ and $(y_2, \pi(y_2))$ connected by a straight edge in $V_1$. Consider the path of the form $S_{x_1}, S_{\pi(\overline{x_1})}, \ldots$ in permutation graph $\Pi$. There are two possible cases:

- Case I: The path is of odd length and terminates in $S_{\pi(y_1)}$ (or $S_{\pi(y_2)}$).
- Case II: The path is of even length and terminates in $S_{x_2}$. (including the case $x_2 = \overline{x_1}$).

Figure 6 illustrates these scenarios. If (I) is true, then $x_1$ and $y_1$ (or $y_2$) are matching inputs and so are $x_2$ and $y_2$ (or $y_1$). If (II) is true, then by symmetry the path in $\Pi$ from $S_{y_1}, S_{\pi(\overline{y_1})}, \ldots$ onwards will also be of even length and terminate in $S_{y_2}$ in which case $x_1$ and $x_2$ can be matched interchangeably with $y_1$ and $y_2$.

Matches for the remaining migrating inputs in $I_0$ and $I_1$ can be defined similarly.

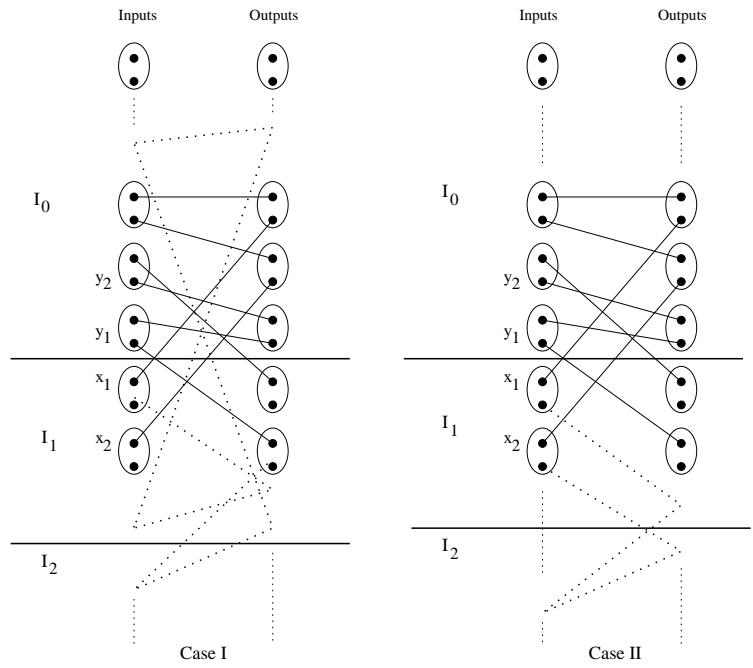

Fig. 11
FINDING MATCHING INPUTS IN $\Pi$.